\documentclass[twoside,a4paper,11pt]{sca}
\usepackage{graphicx}
\usepackage{hyperref}
\usepackage{movie15}

\usepackage{natbib}
\newcommand{\hii}{H~{\sc ii}}
\newcommand{\msinks}{M_\mathrm{sinks}}

\newcommand{\mmax}{M_\mathrm{max}}

\begin{document}
\pagenumbering{arabic}
\pagestyle{myheadings}
\thispagestyle{empty}
{\flushright\includegraphics[width=\textwidth,bb=90 650 520 700]{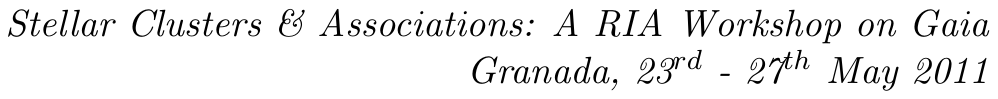}}
\vspace*{0.2cm}
\begin{flushleft}
{\bf {\LARGE
%
Radiative Feedback in Massive Star and Cluster Formation
%
}\\
\vspace*{1cm}
%
Thomas Peters$^{1}$,
Ralf S. Klessen$^{1}$, 
Mordecai-Mark Mac Low$^{2}$,
and 
Robi Banerjee$^{1,3}$
%
}\\
\vspace*{0.5cm}
%
$^{1}$
Zentrum f\"{u}r Astronomie der Universit\"{a}t Heidelberg, Institut f\"{u}r Theoretische Astrophysik, Albert-Ueberle-Str. 2, D-69120 Heidelberg, Germany\\
$^{2}$
Department of Astrophysics, American Museum of Natural History, 79th Street at Central Park West, New York, New York 10024-5192, USA\\
$^{3}$
Hamburger Sternwarte, Gojenbergsweg 112, D-21029 Hamburg, Germany
%
\end{flushleft}
%
\markboth{
Radiative Feedback in Massive Star and Cluster Formation
}{ 
%
Thomas Peters et al.
%
}
\thispagestyle{empty}
\vspace*{0.4cm}
\begin{minipage}[l]{0.09\textwidth}
\ 
\end{minipage}
\begin{minipage}[r]{0.9\textwidth}
\vspace{1cm}
\section*{Abstract}{\small
%
Understanding the origin of high-mass stars is central to modern astrophysics. We shed light
on this problem with simulations using a novel, adaptive-mesh, ray-tracing algorithm. These
simulations consistently follow the gravitational collapse of a massive molecular cloud core,
the subsequent build-up and fragmentation of the accretion disk surrounding the nascent star,
and, for the first time, the interaction between its intense UV radiation field and the infalling
material. We show that ionization feedback can neither stop protostellar mass growth nor
suppress fragmentation. We discuss the effects of feedback by ionizing and non-ionizing radiation
on the evolution of the stellar cluster. The accretion is not limited by radiative feedback but by
the formation of low-mass companions in a process we call ``fragmentation-induced starvation''.
This behavior consistently reproduces the observed relation between the most massive star and
the total mass of stars in a cluster. We show that magnetic fields reduce the star formation rate
and lead to the formation of more massive stars.
%
\normalsize}
\end{minipage}
%
%
%
\section{Introduction \label{intro}}

High-mass stars form in denser and more massive cloud cores
\citep{motteetal08} than their low-mass counter\-parts
\citep{myersetal86}. High densities also
result in local gravitational instabilities in the accretion flow,
resulting in the formation of multiple additional stars
\citep{klesbur00,krattmatz06}.  Young massive stars are almost always
observed to have companions \citep{hohaschik81}, and the number of
their companions significantly exceeds those of low-mass stars
\citep{zinnyork07}.  Such companions influence subsequent accretion
onto the initial star \citep{krumholzetal09}.
Observations show an upper mass limit of about $100\,$M$_{\odot}$. It remains unclear
whether limits on internal stability or termination of accretion by stellar feedback
determines the value of the upper mass limit \citep{zinnyork07}.

The most significant differences between massive star formation and
low-mass star formation seem to be the clustered nature of star
formation in dense accretion flows and the ionization of these flows.
We present the first three-dimensional simulations of the collapse
of a molecular cloud to form a cluster of massive stars
that include ionization feedback by \citet{petersetal10a,petersetal10b,petersetal10c,petersetal11a},
allowing us to study these effects simultaneously.

\section{Numerical Method and Initial Conditions}

We present the first three-dimensional, radiation-(magneto-)hydrodynamical simulations of
massive star formation, taking into account heating by both ionizing and non-ionizing radiation,
using the adaptive-mesh code FLASH \citep{fryxell00}. We propagate the radiation on the adaptive
mesh with our extended version of the hybrid characteristics raytracing method \citep{rijk06,petersetal10a}.
We use sink particles \citep{federrathetal10} to model young
stars. Sink particles are inserted when the Jeans
length of collapsing gas can no longer be resolved on the adaptive
mesh. They continue to accrete any high-density
gas lying within their accretion radius. We use the sink particle mass and accretion rate
to determine the radiation feedback with a prestellar model \citep{petersetal10a}. A detailed
description of our numerical method and the underlying assumptions can be found in \citet{petersetal10a}.

We start our simulations with a $1000\,M_\odot$ molecular cloud having a constant density core
with $\rho = 1.27 \times 10^{-20}\,$ g\,cm$^{-3}$ within a radius of $r = 0.5\,$pc, surrounded by
an $r^{-3 / 2}$ density fall-off out to $r = 1.6\,$pc.
The cloud rotates as a solid body with an angular velocity
$\omega = 1.5 \times 10^{-14}\,$s$^{-1}$. The initial temperature is $T = 30\,$K.
The highest resolution cells on our adaptive mesh have a size of $98\,$AU. Sink particles are
inserted at a cut-off density of $\rho_{\mathrm{crit}} = 7 \times 10^{-16}\,$g\,cm$^{-3}$ and
have an accretion radius of $r_{\mathrm{sink}} = 590\,$AU.

We compare the results of four different simulations. In the first
simulation (Run~A), a dynamical temperature floor is
introduced to suppress secondary fragmentation. Only one sink particle
(representing a massive protostar) is allowed to form. In the second
simulation (Run~B), secondary fragmentation is allowed, and many sink
particles form, representing a group of stars, each contributing to the radiative
feedback. The third
simulation (Run~D) is a control run in which secondary fragmentation
is still allowed, but no radiation feedback is included.
The fourth simulation (Run~E) is a magnetized version of Run~B,
the full stellar group simulation with radiation feedback from all
stars. Run~E includes an initially homogeneous magnetic field along
the rotation axis of the cloud with a magnitude of $10\,\mu$G,
corresponding to a mass-to-flux ratio $(M/\Phi) = 14 (M/\Phi)_{\mathrm{cr}}$ in the central
core at the beginning of the simulation. See \citet{petersetal10c} and
\citet{petersetal11a} for a thorough discussion of these initial conditions.

\section{Fragmentation-Induced Starvation}

\begin{figure}
\center
\includegraphics[scale=0.8]{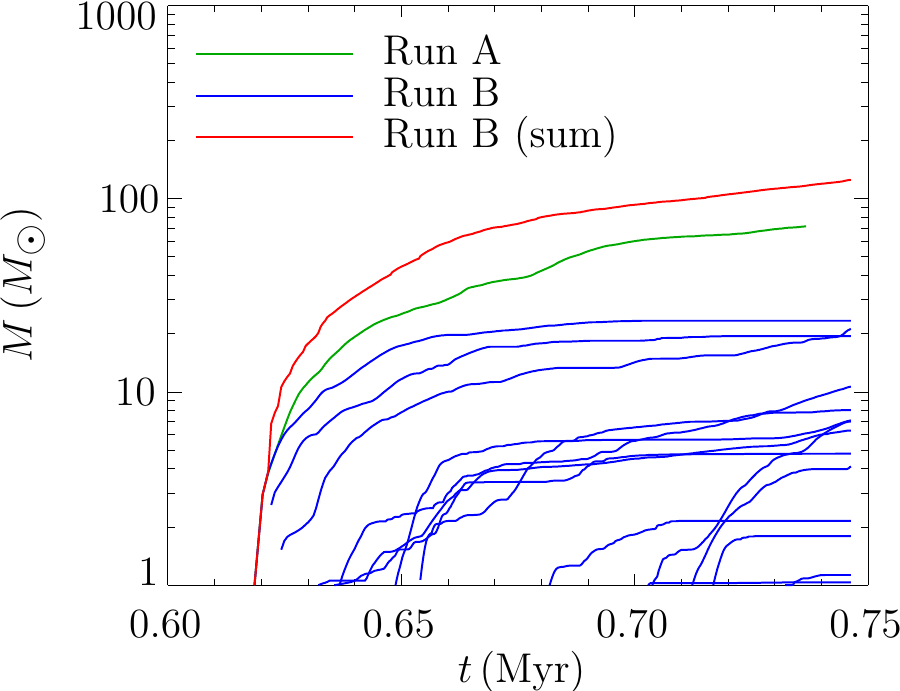} ~
\includegraphics[scale=0.8]{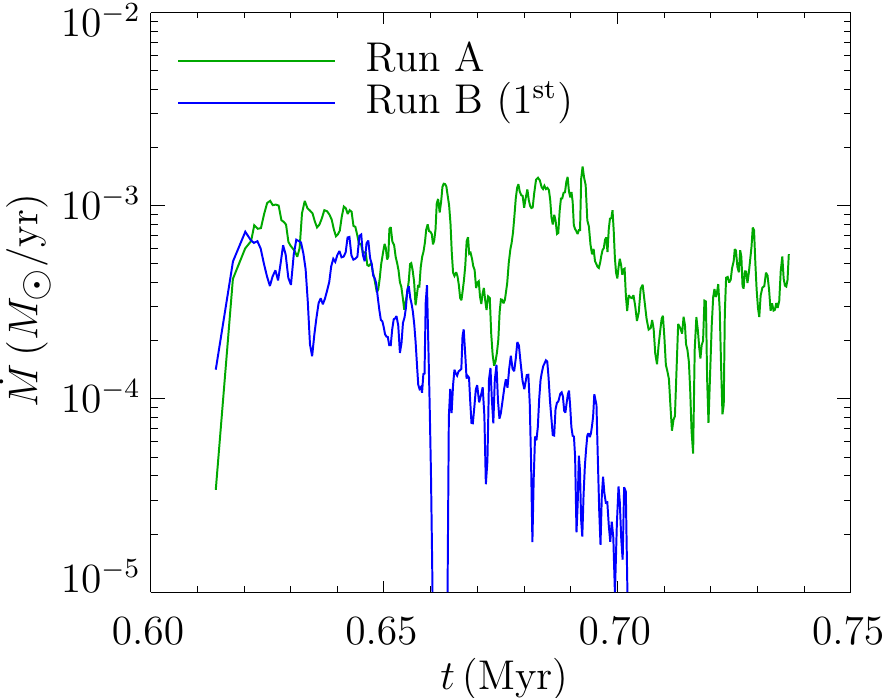} 
\caption{\label{fig2}  Accretion history of the single (Run A) and multiple (Run B)
  sink simulations. Run A was stopped at $72\,$M$_{\odot}$, while no
  sink particle in Run B exceeds $25\,$M$_{\odot}$ over the simulation
  runtime. The left hand plot shows the sink particle masses for Run A
  (green), the individual sink masses of Run B (blue) as well as the
  total mass in sink particles in Run B (red). The right hand plot
  shows the accretion rates of the sink particle in Run A (green) and
  the sink particle which forms first in Run B (blue), which also turns
  out to end up as the most massive at the end of the simulation. While the accretion rate in Run
  A never drops below $10^{-5}\,$M$_{\odot}\,$yr$^{-1}$, accretion
  onto the most massive sink can drop significantly below this value
  and even be stopped totally in Run B.
}
\end{figure}

We first examine the accretion histories of our different models.
Figure~\ref{fig2} shows that in Run~A, with only one sink
particle allowed to form, nothing halts accretion
onto the central protostar. It
continues to grow at an average rate of $\dot{M} \approx 5.9
\times 10^{-4}\,$M$_{\odot}\,$yr$^{-1}$ until we stop the calculation
when the star has reached $72\,$M$_{\odot}$.
The increasingly massive star ionizes the surrounding gas, raising it to high pressure. This
gas breaks out above and below the disk plane, but it cannot halt
mass growth through the disk\footnote{We will hereafter refer to the flattened,
           dense, accretion flow that forms in the midplane of our
           rotating core as a disk.  However, it is not necessarily a
           true Keplerian, viscous, accretion disk, which probably
           only forms within the central few hundred astronomical
           units, unresolved by our models.}
mid-plane.

In contrast, in Run B, where multiple sink particles are allowed to form,
two subsequent sink particles form and begin accreting soon after the first one,
and many more follow within the next $10^{5}\,$yr (see Figure~\ref{fig2}).
By that time the
first sink has accreted $8\,$M$_{\odot}$. Within another $3 \times
10^{5}\,$yr seven further fragments have formed, with masses ranging
from $0.3\,$M$_{\odot}$ to $4.4\,$M$_{\odot}$ while the first three
sink particles have masses between $10\,$M$_{\odot}$ and
$20\,$M$_{\odot}$, all within a radius of $0.1\,$pc from the most
massive object. Accretion by the secondary sinks terminates the mass
growth of the central massive sinks. Material that moves inwards
through the disk driven by gravitational torques accretes
preferentially onto the sinks at larger radii \citep{bate02}. Eventually,
hardly any gas makes it all the way to the center to fall onto the
most massive objects. This fragmentation-induced starvation
prevents any star from reaching a mass $>25\,$M$_{\odot}$ in this case.

Figure~\ref{fig2} also reveals that the most massive stars in the cluster
are those which form first and then keep accreting at a high rate. The most massive stars
in Run~B are the same for all times, but their mass keeps increasing during the formation of the
stellar cluster.

\begin{figure}
\center
\includegraphics[scale=0.8]{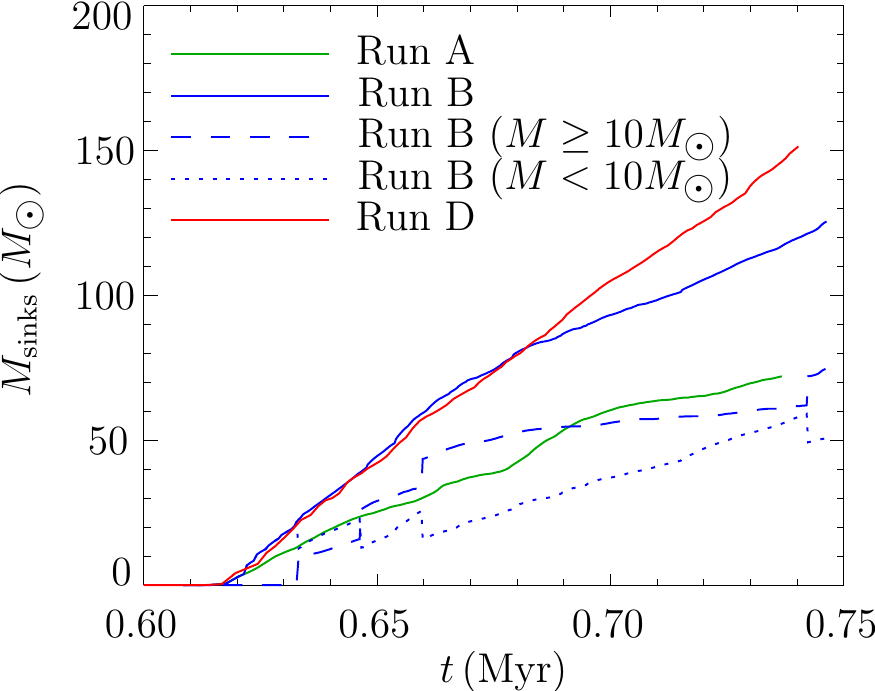} ~
\includegraphics[scale=0.8]{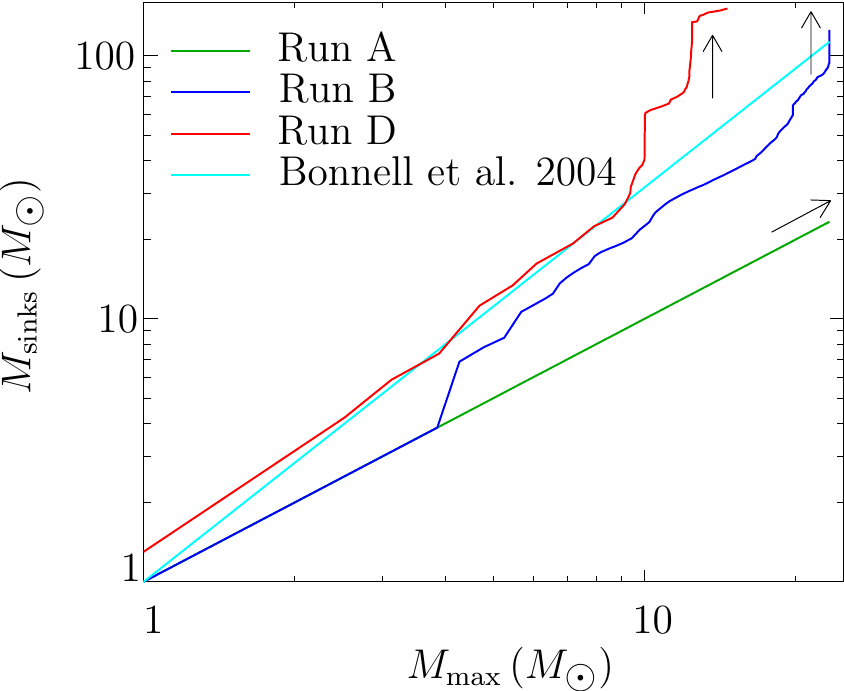} 
\caption{\label{fig3}({\em Left})
Total accretion history of all sink particles combined forming in
Runs A, B, and D. While the heating by non-ionizing radiation does not affect the
total star formation rate, the ionizing radiation appreciably reduces
the total rate at which gas converts into stars once the most massive object has stopped accreting
and its \hii\ region can freely expand. 
The slope of the total accretion history in Run~B goes down because the massive stars (dashed line)
accrete at a decreased rate, while the low-mass stars (dotted line) keep accreting at the same rate.  
({\em Right}) The total mass in sink particles $\msinks$ as a function of the most massive star in the
cluster $\mmax$. We plot the curves for Run~A, Run~B and Run~D as
well as the fit from the competitive accretion simulations by \citet{bonetal04}. 
The turn-off away from the scaling relation (indicated by arrows) is shifted towards higher
masses by radiative feedback. 
}
\end{figure}

The accretion behavior in Run~B contrasts sharply with competitive accretion models
\citep{bonnell01,bonetal04}, which have no mechanism to turn off accretion
onto the most massive stars. In these models, material falls all the way
down to the massive stars sitting in the center of the gravitational potential
which thereby take away the gas from the surrounding low-mass stars. In our fragmentation-induced
starvation scenario, exactly the opposite happens.
Figure~\ref{fig3} illustrates that,
although the accretion rates of the most massive stars ($M \geq
10\,$M$_\odot$) steadily decrease, the low-mass stars ($M <
10\,$M$_\odot$), keep accreting at the same rate.


Competitive accretion models show a correlation between the mass of the most massive
star $\mmax$ and the total cluster mass $\msinks$ during the whole cluster evolution
that is roughly $\mmax \propto \msinks^{2 / 3}$ \citep{bonetal04}. This correlation has
been argued to represent a way to observationally confirm competitive accretion \citep{krumbon07}
and is in fact in good agreement with observations \citep{weidkroup06,weidetal10}.
However, we find that our simulations also reproduce the observed relation between $\mmax$ and $\msinks$.

Figure~\ref{fig3} shows $\msinks$ as function of $\mmax$ for Run~A, Run~B and Run~D,
and the relation $\mmax = 0.39 \msinks^{2/3}$, which was found by \citet{bonetal04}
as a fit to their simulation data. Over the whole cluster evolution, the curve for Run~D lies
above this fit, while the curve for Run~B always lies below it. The fit agrees
with our simulation data as well as it does to that of \citet{bonetal04}.
This indicates that the
scaling is not unique to competitive accretion, but can also be found with the fragmentation-induced
starvation scenario and hence cannot be used as an observational confirmation of competitive
accretion models.


\section{Effects of Radiative Feedback and Magnetic Fields}

\begin{figure}
\center
\includegraphics[scale=0.8]{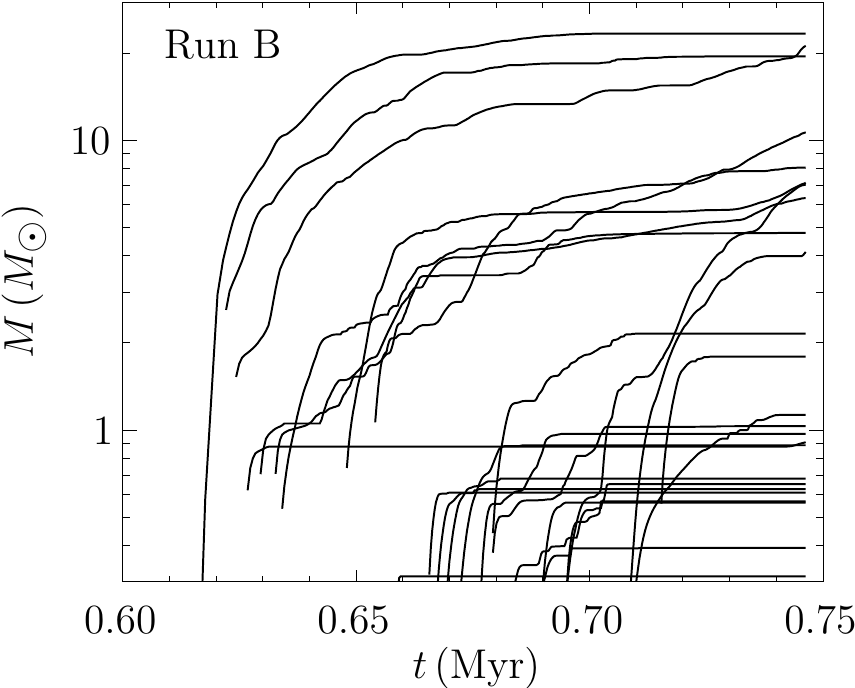} ~
\includegraphics[scale=0.8]{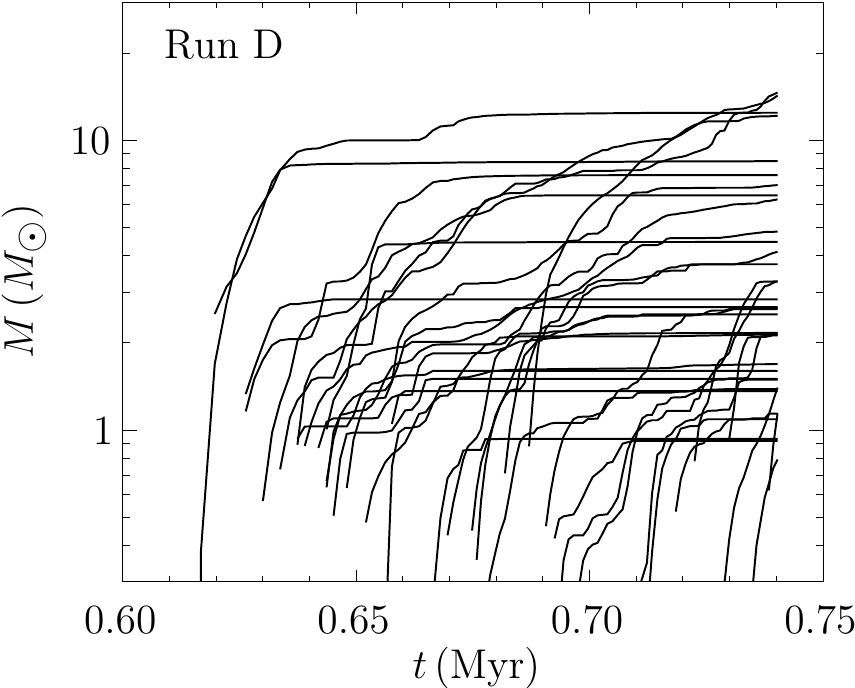} 
\caption{\label{fig4} Individual accretion histories for Run~B and Run~D. The figure shows the stellar
masses as function of time for all sink particles that form in Run~B ({\em left}) and Run~D ({\em right}).
Because the Jeans mass is lower without radiative feedback (Run~D), many
more sink particles form in Run~D than in Run~B, and the mass of the most massive stars is also lower.
}
\end{figure}

Figure~\ref{fig4} shows the individual accretion histories of each of the sink
particles in Run~B and Run~D. Radiative heating cannot prevent disk fragmentation but
raises the local Jeans mass. Hence, many fewer stars form in Run~B than
in Run~D, and the mass of the most massive stars in Run~B is higher. It is also evident
from the figure that star formation is much more intermittent in the case with radiation feedback
(Run~B). The reason for
this behavior is that the star formation process is controlled by the local Jeans mass,
which depends to a large degree on how the filaments in the disk shield the radiation.
Shielding can lower the Jeans mass temporarily
and thereby allow gravitational collapse that would not have occurred otherwise.

Since the accretion heating raises the Jeans mass and
length in Run~B,
the total number of sink particles is higher in Run~D than in Run~B,
      and the
stars in Run~D generally reach a lower mass than in
Run~B. These two effects
cancel out
to lead to the same overall star formation
          rate for some time (see Figure~\ref{fig3}).
At one point in the evolution, however, also
the total accretion rate of Run~B drops below that of Run~D.  At time
$t \approx 0.68\,$Myr the accretion flow around the most massive star
has attenuated below the value required to trap the \hii\ region. It
is able to break out and affect a significant fraction of the disk
area. A comparison with the mass growth of Run~D clearly shows that
there is still enough gas available to continue constant cluster
growth for another $50\,$kyr or longer, but the gas can no longer
collapse in Run~B.  Instead, it is swept up in a shell surrounding the
expanding \hii\ region.

It is also notable that the expanding ionization front around the most
massive stars does not trigger any secondary star formation,
which suggests that triggered star formation \citep{elmelad77} may not be as efficient as expected,
at least on the scales considered here.

\begin{figure}
\center
\includegraphics[scale=0.8]{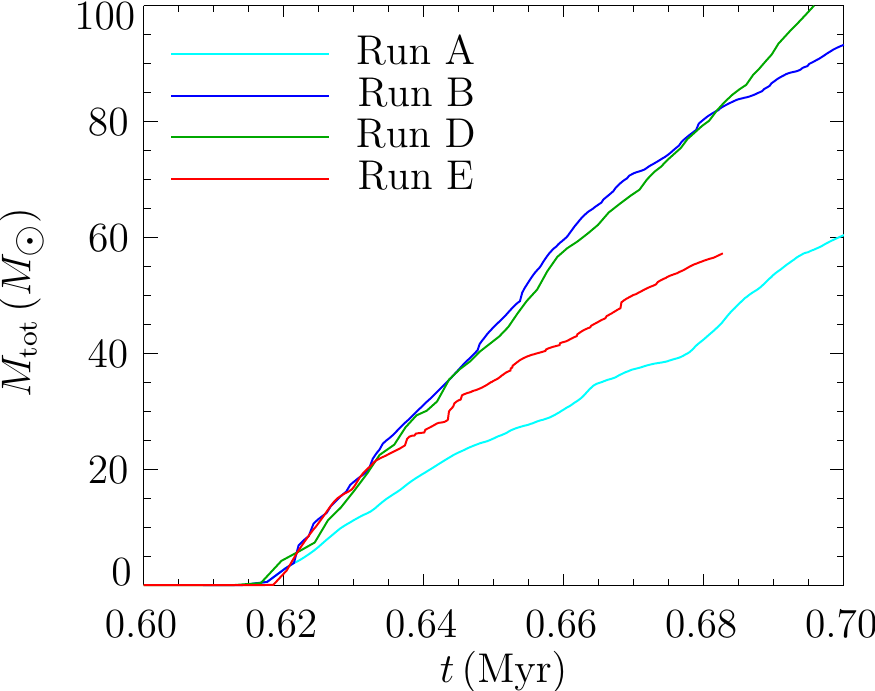} ~
\includegraphics[scale=0.8]{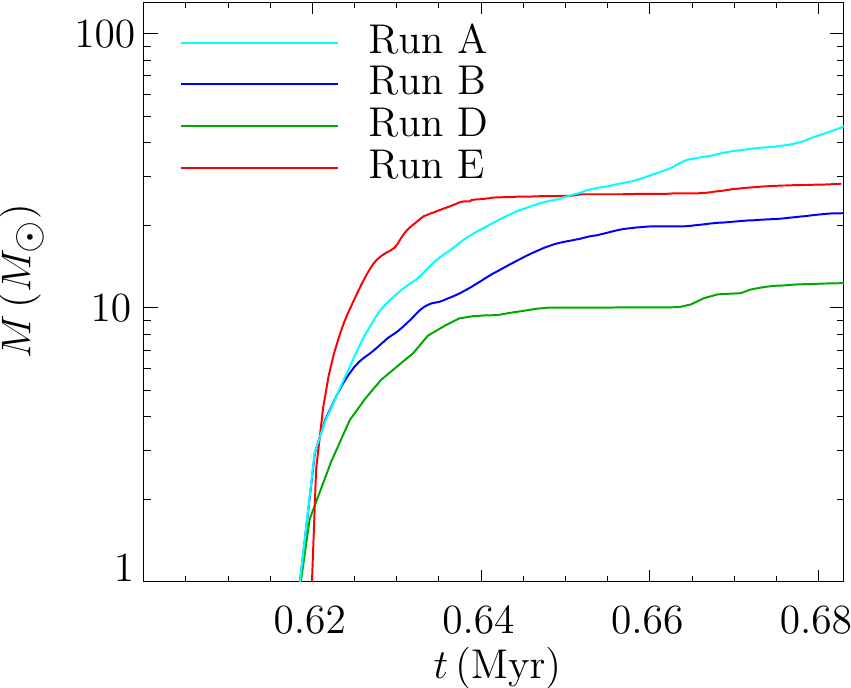} 
\caption{\label{fig5} ({\em Left}) Total accretion history of Run~A, Run~B, Run~D and Run~E. While the total accretion rates
of Run~B and Run~D agree until ionization feedback stops massive star formation in Run~B, the
total accretion rate in Run~E already starts to decline after $20\,$kyr. The magnetic field in Run~E
additionally supports the gas against collapse, reducing the total accretion rate.
({\em Right}) Protostellar masses of the first sink particles in Run~A, Run~B, Run~D and Run~E.
Radiative heating (Run~B) and presence of magnetic fields (Run~E) increase the final masses of the massive stars.
The largest part of the additional mass accretion in Run~E compared to Run~B is due to the
stronger initial accretion phase.
}
\end{figure}

The total accretion histories of all sink particles combined in each of
the four simulations are contrasted in Figure~\ref{fig5}.
For the first $20\,$kyr, the total accretion rate of Run~B, Run~D and Run~E is nearly identical.
The accretion rate in these multiple sink simulations is generally higher than in the single sink
Run~A since the large group of sinks accretes from a large volume, without needing to
rely on outward angular momentum transport to deliver material to the direct feeding zone
of the central sink. In the magnetized Run~E, no secondary sink particles form during the first $20\,$kyr,
so that the increased accretion rate
in this phase is due to the additional angular momentum transport
performed by the magnetic field.
After the initial $20\,$kyr, the total accretion rate in Run~E falls
below Run~B and the control Run~D, deviating
increasingly with time, but always staying above the accretion rate in Run~A.
The accretion histories of Run~B and Run~D only start to separate at
relatively late time when the ionizing radiation begins to terminate accretion onto the most massive
sinks in Run~B (see the discussion in \citealt{petersetal10c}), but the magnetic field additionally
reduces the rate at which gas collapses in Run~E. Thus, the total accretion
rate with magnetic fields is lower than without magnetic fields.

The central star in Run~E can grow to larger masses because in the initial phase
fragmentation is delayed by magnetic support. This allows the central
object to maintain a high accretion rate for a longer time.
Nevertheless, the accretion rate of the massive star
in Run~E drops significantly when secondary sink particles form, since
they form in a dense ring around the central massive sink and to some
extent starve it of material, even though they never cut off accretion
entirely.

The stronger initial accretion phase in Run~E compared to
Run~B yields the main contribution to the larger final mass of the
massive sink. The magnetic field very efficiently redistributes
angular momentum, resulting in an increased radial mass flux through the
high-density equatorial plane. Consequently, the initial accretion
rate in the magnetized simulation (Run~E) even lies considerably above
the non-magnetized single sink calculation (Run~A).

\section{Conclusions}

We have reviewed some of the results of our recent radiation-(magneto-)hydrodynamical simulations of
massive star formation \citep{petersetal10a,petersetal10b,petersetal10c,petersetal11a},
which for the first time simultaneously include the effect of heating
by both ionizing and non-ionizing radiation. We find that ionization feedback is
unable to stop protostellar mass growth. Instead, the mass of the most massive stars is limited
by the formation of lower-mass companions in their gravitationally unstable accretion flow
in a process we call fragmentation-induced starvation. Our numerical model reproduces the
observed relation between the most massive star in a cluster and the total cluster mass.
We find that heating by non-ionizing radiation decreases the total number of stars formed
and increases the mass of the most massive stars, but does not change the total star formation rate.
The star formation rate is reduced by ionizing radiation once the \hii\ regions can steadily expand.
Even initially very weak magnetic fields can markedly reduce the star formation rate by
providing additional support against gravitational collapse and lead to the formation of more
massive stars via magnetic braking of the accretion flow.

\vspace{1cm}

\small


%
%
%
%
%
\end{document}